\documentclass[conference]{IEEEtran}
\IEEEoverridecommandlockouts
\usepackage{amsmath,amssymb,amsfonts}
\usepackage{listings}
\usepackage{algorithm}
\usepackage{algorithmicx}
\usepackage[noend]{algpseudocode}
\usepackage{graphicx}
\usepackage{textcomp}
\usepackage{xcolor}
\usepackage{tikz}
\usetikzlibrary{shapes,positioning,calc}
\usepackage[hyphens]{url}
\usepackage[unicode, hidelinks]{hyperref}
\usepackage[hyphenbreaks]{breakurl}
\usepackage[style=ieee, bibencoding=utf8, citestyle=numeric-comp]{biblatex}
\def\BibTeX{{\rm B\kern-.05em{\sc i\kern-.025em b}\kern-.08em
    T\kern-.1667em\lower.7ex\hbox{E}\kern-.125emX}}
\usepackage{booktabs}
\usepackage{colortbl}
\usepackage{multirow}
\usepackage{flushend}
\usepackage[inline]{enumitem}
\usepackage{eqparbox}
\newdimen{\algindent}
\algnewcommand\LeftComment[2]{%
\hspace{#1\algindent}$\triangleright$ \eqparbox{COMMENT}{#2} \hfill %
}
\algnewcommand{\algorithmicand}{\textbf{ and }}
\algnewcommand{\algorithmicor}{\textbf{ or }}
\algnewcommand{\OR}{\algorithmicor}
\algnewcommand{\AND}{\algorithmicand}

\begin{document}

\title{Symbolic Security Predicates: \\
Hunt Program Weaknesses
\thanks{This work was supported by RFBR grant 20-07-00921 A.}}

\author{
\IEEEauthorblockN{
  Alexey Vishnyakov\IEEEauthorrefmark{1},
  Vlada Logunova\IEEEauthorrefmark{1}\IEEEauthorrefmark{3},
  Eli Kobrin\IEEEauthorrefmark{1}\IEEEauthorrefmark{2},
  Daniil Kuts\IEEEauthorrefmark{1},
  Darya Parygina\IEEEauthorrefmark{1}\IEEEauthorrefmark{2} and
  Andrey Fedotov\IEEEauthorrefmark{1}
}
\IEEEauthorblockA{
  \IEEEauthorrefmark{1}Ivannikov Institute for System Programming of the RAS
}
\IEEEauthorblockA{
  \IEEEauthorrefmark{2}Lomonosov Moscow State University
}
\IEEEauthorblockA{
  \IEEEauthorrefmark{3}Moscow Institute of Physics and Technology
}
Moscow, Russia \\
\{vishnya, vlada, kobrineli, kutz, pa\_darochek, fedotoff\}@ispras.ru
}

\maketitle

\begin{tikzpicture}[remember picture, overlay]
\node at ($(current page.south) + (0,0.65in)$) {
\begin{minipage}{\textwidth} \footnotesize
  Vishnyakov A., Logunova V., Kobrin E., Kuts D., Parygina D., Fedotov A.
  Symbolic Security Predicates: Hunt Program Weaknesses. 2021 Ivannikov ISPRAS
  Open Conference (ISPRAS), IEEE, 2021, pp. 76-85. DOI:
  \href{https://www.doi.org/10.1109/ISPRAS53967.2021.00016}{10.1109/ISPRAS53967.2021.00016}.

  \copyright~2021 IEEE. Personal use of this material is permitted. Permission
  from IEEE must be obtained for all other uses, in any current or future media,
  including reprinting/republishing this material for advertising or promotional
  purposes, creating new collective works, for resale or redistribution to
  servers or lists, or reuse of any copyrighted component of this work in other
  works.
\end{minipage}
};
\end{tikzpicture}

\begin{abstract}
Dynamic symbolic execution (DSE) is a powerful method for path exploration
during hybrid fuzzing and automatic bug detection. We propose security
predicates to effectively detect undefined behavior and memory access violation
errors. Initially, we symbolically execute program on paths that don't trigger
any errors (hybrid fuzzing may explore these paths). Then we construct a
symbolic security predicate to verify some error condition. Thus, we may change
the program data flow to entail null pointer dereference, division by zero,
out-of-bounds access, or integer overflow weaknesses. Unlike static analysis,
dynamic symbolic execution does not only report errors but also generates new
input data to reproduce them. Furthermore, we introduce function semantics
modeling for common C/C++ standard library functions. We aim to model the
control flow inside a function with a single symbolic formula. This assists bug
detection, speeds up path exploration, and overcomes overconstraints in path
predicate. We implement the proposed techniques in our dynamic symbolic
execution tool Sydr. Thus, we utilize powerful methods from Sydr such as path
predicate slicing that eliminates irrelevant constraints.

We present Juliet Dynamic to measure dynamic bug detection tools accuracy. The
testing system also verifies that generated inputs trigger sanitizers. We
evaluate Sydr accuracy for 11 CWEs from Juliet test suite. Sydr shows 95.59\%
overall accuracy. We make Sydr evaluation artifacts publicly available to
facilitate results reproducibility.
\end{abstract}


\section{Introduction}

Modern software is rapidly developing. Novel code inevitably brings new bugs and
weaknesses~\cite{cwe}. More and more industrial companies follow security
development lifecycle~\cite{howard06, iso08, gost16} to improve application
quality and defend it from malicious attacks. Coverage-guided
fuzzing~\cite{serebryany16, sargsyan19, fioraldi20} is continuously applied to
detect program crashes during development process. Advanced hybrid fuzzing
benefits from dynamic symbolic execution (DSE)~\cite{king76, saudel15,
stephens16, yun18, baldoni18, vishnyakov20, poeplau20, poeplau21,
borzacchiello21, kuts21}. DSE discovers complex program states that are hardly
reachable via simple fuzzing.

The main focus of this paper is the promising bug-driven hybrid testing
approach~\cite{chen20, osterlund20}. We propose automatic error detection for
undefined behaviour and
memory access violation errors. In particular, we utilize dynamic symbolic
execution to generate program input data that trigger integer
overflow~\cite{wang09, demidov17, chen20}, out-of-bounds access~\cite{chen20},
etc. These are the most widespread and dangerous program
weaknesses~\cite{cwe-top25}. We symbolically execute binary code because it's
more portable than building from varying programming languages. The fuzzing
setup may be the following. We build a target program with
sanitizers~\cite{song19} for fuzzer and without them for our dynamic symbolic
execution tool Sydr~\cite{vishnyakov20}. Sydr explores new program states via
branch inversion. Moreover, it generates new inputs that trigger errors. All
inputs are passed to fuzzer that verifies inputs on sanitizers. Thus, fuzzer
takes inputs from Sydr that increase code coverage and verifies whether any
inputs crash program on sanitizers.

This paper makes the following contributions:
\begin{itemize}
  \item We introduce symbolic function semantics for common C/C++ standard
    library functions. We aim to model the control flow inside a library
    function with a single symbolic formula. Thus, we can explore more program
    states from one concrete execution path. This speeds up program paths
    exploration and assists bug detection.
  \item We propose security predicates that allow us to effectively detect
    undefined behavior and memory access violation errors on initially valid
    paths. Moreover, dynamic symbolic execution produces inputs that trigger
    such errors. These inputs may later be verified on sanitizers~\cite{song19}.
  \item We present Juliet Dynamic~\cite{juliet-dynamic} to measure dynamic bug
    detection tools accuracy on Juliet test suite~\cite{juliet}. We evaluate
    accuracy of our security predicates implementation in
    Sydr~\cite{vishnyakov20} for 11 CWEs~\cite{cwe} from Juliet. Sydr shows
    95.59\% overall accuracy.
\end{itemize}

The paper is organized as follows. Section~\ref{sec:func} discusses function
semantics modeling during symbolic execution. Section~\ref{sec:security}
presents null pointer dereference, division by zero, out-of-bounds access, and
integer overflow security predicates. Section~\ref{sec:implementation} sheds
light on implementation details. Section~\ref{sec:evaluation} contains function
semantics evaluation, Juliet Dynamic~\cite{juliet-dynamic} description, and Sydr
security predicates evaluation on Juliet test suite~\cite{juliet}.
Section~\ref{sec:conclusion} concludes the paper.

\section{Function Semantics}
\label{sec:func}

Performance is crucial in dynamic symbolic execution. Plenty of optimizations
are implemented to decrease time required for analysis. It's worth to reduce
instrumentation where it is possible since the instrumented process is always
slower than native execution in orders of magnitude.

There is no need to symbolically execute every internal instruction in libc
functions with standard specified semantics. Moreover, cutting-edge concolic
executors adapting such approach benefit from diminishing overconstrainting of
symbolic state~\cite{chipounov12, shoshitaishvili16, poeplau20,
borzacchiello21}. For instance, both uppercase and lowercase characters are
permissible as input data for \texttt{tolower} what can be expressed in the
following if-then-else (\emph{ite}) formula for input symbol \emph{ch}:
$$ite(ch - \texttt{'A'} < 26,\  ch - (\texttt{'A'} - \texttt{'a'}),\  ch)$$
However, relying on concrete execution trace inside this function during
analysis ends up in overconstrainting to letter case from concrete execution.

Instead, we propose function semantic models which can incorporate more symbolic
states in a single constructed formula and speed up the execution. Similar
techniques are applied in a range of notable DSE tools. In our view, there are
four main directions in function semantics modeling which prevail or can be
combined in a certain tool:
\begin{enumerate}
  \item KLEE~\cite{cadar08} replaces libc functions with their simpler versions
    precompiled to LLVM IR. This aims to curtail extra components compilation
    and simplify the analyzed code. This method doesn't involve symbolic
    modeling of entire fucntion.
  \item The function instrumentation is replaced with path
    constraints reproducing concrete execution trace~\cite{poeplau20}.
  \item The function is modeled with a single branch for representative
    states~\cite{borzacchiello21}, e.g. strings are equal in \texttt{strcmp}.
  \item Our approach generally focuses on construction of symbolically computed
    expression for return value~\cite{chipounov12, shoshitaishvili16}.
\end{enumerate}

We handle function semantics similar to Fuzzolic analysis
modes~\cite{borzacchiello21} except that we always perform concrete execution
via dynamic binary instrumentation~\cite{bruening04} to update concrete state.
Symbolic execution starts from the first symbolic input read. We divide symbolic
function models in several groups. Some of them move around already existing
symbolic expressions in memory and create necessary path constraints while
others just skip symbolic execution. The more complicated semantics deal with
comparisons and string to integer conversions that require construction of
symbolic formulas representing return values. It is worth noting that Sydr
always synchronizes concrete and symbolic states. The symbolic state is
concretized if two states mismatch. Thus, we should mind that symbolic formulas
should comply with the concrete state.

\subsection{No Symbolic Computation}

This section comprises functions with trivial side-effects. We handle them to
increase performance and simplify the symbolic formulas.

\subsubsection{Dynamic Memory}

Heap allocation functions (\texttt{malloc}, \texttt{calloc}, \texttt{realloc},
\texttt{free}) often operate on symbolic data. We execute such functions
concretely and maintain symbolic memory state consistency. Thus, we are able to
radically reduce the number of overconstraint branch conditions in the path
predicate. In practice, skipping symbolic execution of internal instructions for
these functions provides significant increase in performance.

\subsubsection{Data Movement}

Symbolic model of copying operations (like \texttt{strcpy}, \texttt{memcpy},
\texttt{memmove}) reassigns symbolic expressions without modification.
Nevertheless, we should consider the null-terminator position as it may spawn
new branches.

\subsubsection{Printing Omission}

Functions which write program log to standard output are not of interest for
symbolic execution process~\cite{borzacchiello21}. Consequently, they can be
skipped in favour of performance improvement. Moreover, this might decrease the
number of irrelevant constraints for the printed data.

\subsection{String Comparison}

We utilize Triton framework~\cite{saudel15} to construct formulas for return
values of functions \texttt{memchr}, \texttt{strlen}, \texttt{strcmp},
\texttt{memcmp}, \texttt{strstr}, and etc. There are two basic semantics in this
group:
\begin{enumerate*}
    \item character search in string or memory area (\`a la \texttt{memchr}),
    \item lexicographical comparison of two strings or memory areas (\`a la
      \texttt{memcmp}).
\end{enumerate*}
The \texttt{strstr} function can be considered as \texttt{strchr} with extended
character argument.

\subsubsection{\`a la \texttt{memchr}}

Formula for \texttt{memchr}-like functions should express pointer to the
character in string or null-pointer when it cannot be found. Both string and
character may be symbolic. We compute the character position via sum of
if-then-else (\emph{ite}) expressions that evaluate to 1 or 0. Here is the
formula for \texttt{memchr(ptr, ch, count)} return value:
\begin{align*}
\begin{aligned}
ite\Bigg(&\bigwedge_{k=0}^{count-1} ptr[k] \neq ch,\ 0, \\
&ptr\ + \sum_{i=0}^{count-1} ite\Bigg(\bigwedge_{k=0}^{i} ptr[k] \neq ch, 1, 0\Bigg)
\Bigg)
\end{aligned}
\end{align*}
The first \emph{ite}-condition checks if the character is present in string and
sets the pointer to null otherwise. In the second part every summand increments
the pointer until the required character is discovered. It is worth noting that
\texttt{strlen(str)} is similar to \texttt{memchr(str, 0, length + 1)}.

When handling strings (e.g. \texttt{strchr}) the formula is burdened with extra
conditions caused by null-terminator consideration. In case of \texttt{strstr},
character search $ptr[k] \neq ch$ is replaced with appropriate conjuctions for
substring comparisons.

\subsubsection{\`a la \texttt{memcmp}}

The task of lexicographical comparison (\texttt{memcmp}, \texttt{strcmp},
\texttt{strncmp}, etc.) can be formulated as mismatch search and difference evaluation. The
\texttt{memcmp(lhs, rhs, count)} formula is the following:
\begin{align*}
\begin{aligned}
lhs[0] &- rhs[0]\ + \\
\sum_{i=1}^{count-1} (lhs[i] &- rhs[i]) *
ite\left(\bigwedge_{k=0}^{i-1} lhs[k] = rhs[k], 1, 0\right)
\end{aligned}
\end{align*}
The formula is designed so that only one summand might be meaningful and others
are equal to zero. We make the conjunction of all previous character pairs
equality to identify the first pair of differing symbols. In string comparison
it extra checks that no null byte pair was discovered before.

The mentioned above function semantics specify the returned value sign but
cannot guarantee concrete and symbolic states consistency. For example, in
32-bit Linux the return values set is $\{1, 0, -1\}$ while x86\_64 uses the
characters difference. For that reason, the built expression should be
additionally tied to current return value to avoid concretization.

\subsection{String to Integer Conversion}

The main focus of this group is on \texttt{strtol}, \texttt{strtoul},
\texttt{strtoll}, and etc. These functions are also called inside other
functions like \texttt{atoi} and \texttt{scanf("\%d", \&x)}. In addition,
similar semantics are applied for \texttt{std::cin} reading of integer types.

The byte string parsed by \texttt{strtol} may have many configurations due to
optional components and can be modeled by splitting into several parts: \\
\texttt{[whitespaces]\textit{[sign][prefix] [valid~symbols]}[invalid~symbols]nullbyte}

The set of valid symbols is defined by the \emph{base} argument and
can consist of digits and latin characters. When the \emph{base} is null,
decimal numeral system is chosen by default or it might be learned from octal
and hexadecimal prefixes (0 and 0x/0X respectively). The string gets through
preliminary parsing to extract parts relevant to conversion
(\texttt{\textit{italic}}). The same structure is preserved during new inputs
generation.

\begin{align}
\label{eq:strtol}
&\pm(c_n c_{n-1} ... c_1 c_0)_b \longrightarrow x \\
&\begin{aligned}\label{eq:chrtod}
a_k = ite(c_k\geq\texttt{'0'} \land \ c_k\leq\texttt{'9'} \land \ c_k < \texttt{'0'} + b, \\
c_k - \texttt{'0'}, \\
ite(c_k \geq \texttt{'a'} \land \ c_k < \texttt{'a'} + b - 10, \\
c_k - \texttt{'a'} + 10,\ c_k - \texttt{'A'} + 10))
\end{aligned} \\
\label{eq:base}
&|x| = \sum_{k=0}^{n} a_k b^k,\ x = ite(sign = \texttt{'-'}, -|x|, |x|) \\
&\begin{aligned}\label{eq:constraints}
(c_k \geq \texttt{'0'} \land\ c_k \leq \texttt{'9'} \land\ c_k < \texttt{'0'} + b&)\ \lor \\
(c_k \geq \texttt{'a'} \land c_k < \texttt{'a'} + b - 10&)\ \lor \\
(c_k \geq \texttt{'A'} \land c_k < \texttt{'A'} + b - 10&)
\end{aligned}
\end{align}

In order to implement the symbolic string to integer
conversion~(\ref{eq:strtol}) we constrain every character~(\ref{eq:constraints})
to assure they lay within valid symbols according to base \emph{b} determined
from argument value or parsing. The formula~(\ref{eq:chrtod}) depicts how we
compute digit from character, i.e. classical digit symbol or alphabet letter in
lower/upper case. The whole number computation result can be observed in the
formula~(\ref{eq:base}).

The proposed model concretizes \emph{base} argument because symbolic base
produces varying scenarios with corner cases that require lots of complex
symbolic formulas. So, we don't consider symbolic \emph{base} for the sake of
model simplicity. Thus, we should also constrain the base prefix. When the
\emph{base} argument is zero, the leading zeros would cause decimal base
distortion into octal. For instance, the initial input 4567 defines $base = 10$
and new input 0123 shouldn't be generated for number $123_{10}$ as it will be
naturally identified as $123_{8}$. The above corner case can be solved by
replacing leading zero digits with spaces. Furthermore, additional constraints
are constructed to make sure the space symbols wouldn't appear between digits.
However, this situation wouldn't occur if the \emph{base} argument is a concrete
non-zero value.

The obtained formula~(\ref{eq:base}) may overflow, i.e. the computed value won't
fit in return type (\texttt{long} for \texttt{strtol}). Therefore, we
compute~(\ref{eq:base}) in a bit vector of twice bigger size than return type.
Afterwards, we constrain the computed value to fit in return type, e.g.
$LONG\_MIN \leq x \leq LONG\_MAX$.

\section{Security Predicates}
\label{sec:security}

Dynamic symbolic execution is a powerful method to discover new program paths
during hybrid fuzzing~\cite{yun18, vishnyakov20, poeplau20, poeplau21,
borzacchiello21, kuts21}. Furthermore, we show that it can be effective to
detect bugs on valid paths. We symbolically execute a program with input that
doesn't lead to crash. If instruction or function operates on symbolic data,
then we construct security predicates that check for undefined behavior and
memory access violation. \textit{Security predicate} for some error type
(weakness) is a Boolean predicate that holds true iff the instruction (or
function) triggers an error. Further, we conjunct a security predicate with
sliced branch constraints from the path predicate, i.e. constraints over
symbolic variables that are relevant to variables in security
predicate~\cite{vishnyakov20}. Thus, we guarantee that program reaches the
examined instruction. Finally, we call solver to resolve the constructed
constraint. If it is satisfiable, then Sydr reports an error and generates new
program input from obtained model. The generated input will reproduce the error
in analyzed program.

In particular, Sydr can detect division by zero and null pointer dereference
errors. Security predicates for them are equality of the divisor and memory
dereference address respectively to zero. Let's consider how Sydr generates
inputs that trigger null pointer dereference. The following example
contains possible null pointer dereference in line~6:
\begin{lstlisting}[language=C, basicstyle=\small\ttfamily, numbers=left,
                   xleftmargin=2em,
                   caption={NULL pointer dereference (Linux x64).},
                   captionpos=b,
                   label=lst:null]
char a[] = {'1', '2', '3', '4', '5'};
int main() {
    long i;
    scanf("%ld", &i);
    if (i >= 5) return 1;
    printf("%c\n", a[i]);
}
\end{lstlisting}
We run this example under Sydr with +0000000000000000001 as standard input that
is enough to generate any number that can fit in 64-bit signed long integer
type. This comes from limitation of dynamic symbolic execution approach. DSE
cannot increase initial concrete input size, otherwise it won't interpret
missing instructions operating on the additional input bytes and symbolic model
will be unsound. The symbolic input is obtained via \texttt{read} system call
inside \texttt{scanf} function. Sydr creates a symbolic variable for each
inputted byte. Starting from this point Sydr interprets all assembly
instructions that operate on symbolic values (updates symbolic registers and
memory states). The necessary concrete values are gathered from
DynamoRIO~\cite{bruening04}. Then \texttt{scanf} calls
\texttt{\_\_strtoll\_internal} to convert input string to integer. Sydr wraps
\texttt{strtoll} and constructs corresponding formula for string to integer
conversion. Later this formula is compared with size of array $a$ in line 5 and
$i < 5$ constraint is pushed to path predicate. In line 6 the index value $i$
is symbolic. So, we verify whether symbolic address $a + i$ can be zero. We
conjunct the null dereference security predicate with the path predicate. The
resulting predicate $i < 5~\land~a + i = 0$ is satisfiable because line 5 does
not consider that integer may be negative. Thus, Sydr reports null pointer
dereference on \texttt{movzx eax, byte ptr [rdx + rax]} instruction, where \texttt{rax}
is array $a$ base address and \texttt{rdx} is index $i$. Moreover, Sydr stores new
stdin value -0000000000006295616 that triggers zero dereference.

Null pointer dereference security predicate is able to generate input computing
to actual zero address when binary is complied without ASLR. Though it is still
fast and easy solvable checker that can simply generate inputs that cause
segmentation faults even when ASLR is enabled.

It is worth noting that some zero dereferences cannot be found with security
predicates. They should be located with simple path discovery via symbolic
execution or fuzzing. For instance, pointer in listing below isn't symbolic but
can be null:
\begin{lstlisting}[language=C, basicstyle=\small\ttfamily, numbers=left,
                   xleftmargin=2em]
int *p = 0;
if (some_condition) p = malloc(4);
*p = 3;
\end{lstlisting}

The main advantage of bug searching with DSE is that it produces inputs to
verify errors. However, true positive rate is not 100\%. That's why we build two
target binaries. First one is with sanitizers for fuzzing. The second one is
without sanitizers for DSE since we don't want to symbolically interpret
instrumentation code. In this approach fuzzer verifies whether generated inputs
from security predicates crash on sanitizers.

\subsection{Out-of-bounds Access}

Out-of-bounds access is the most dangerous and widespread program
error~\cite{cwe-top25}. We build security predicate for each memory access at
symbolic address $sym\_addr$ (that depends on user input) to detect such errors.
Firstly, we determine memory buffer bounds $[lower\_bound, upper\_bound)$ and
create predicate that is true when symbolic address is outside of these bounds
$sym\_addr < lower\_bound \lor sym\_addr >= upper\_bound$. Afterwards, we
conjunct this predicate with sliced~\cite{vishnyakov20} path constraints. If
solver returns a model for constructed predicate, then Sydr reports an
out-of-bounds error and generates a corresponding input, i.e. original input
with bytes replaced with ones from the model. It is worth noting that both
bounds cannot be always determined in binary code. Thus, we try to point symbolic
address outside (below or above) the single detected bound. For instance, in
Listing~\ref{lst:null} out-of-bounds is also feasible but Sydr cannot detect an
upper bound for global array $a$. However, Sydr is able to heuristically
retrieve the array base address from symbolic address expression \texttt{[rdx +
rax]}, where \texttt{rax} is concrete array base address and \texttt{rdx} is
symbolic index. Sydr assumes the concrete part \texttt{rax} to be lower bound.
Thus, it generates input that triggers access below the lower bound (e.g. -1).

We maintain shadow heap and stack in order to determine symbolic address bounds.
Sydr wraps all dynamic memory management functions (\texttt{malloc},
\texttt{calloc}, \texttt{realloc}, \texttt{free}, etc.) and appropriately
updates the shadow heap that holds all allocated memory buffer bounds. On each
call instruction Sydr pushes return address location (stack pointer value) to
shadow stack. And on all call and return instructions Sydr pops elements from
shadow stack according to current stack pointer value.

When instruction accesses memory at symbolic address Sydr detects corresponding
buffer bounds the following way. If current concrete address value is in shadow
heap, then both bounds are retrieved from it. When address points to stack, the
closest return address location (from shadow stack) above concrete address is
the upper bound. The lower bound is computed heuristically from the concrete
part of symbolic address formula. The main idea behind this approach is to sum
up the concrete parts of the formula. However, it considers some corner cases.
For instance, we should distinguish \texttt{a[i - 0x20]} from array on stack
\texttt{[ebp - 0x20]}. We use the same heuristics to detect global array lower
bound.

Moreover, Sydr wraps memory copy functions (\texttt{memcpy}, \texttt{memmove},
\texttt{memset}, \texttt{strncpy}, etc.) to detect buffer overflows. If memory
copy size function argument is symbolic, Sydr tries to make it exceed the upper
bound.

Before solving the security predicate Sydr conjuncts it with strong precondition
to make error most likely cause a crash, i.e. overwrite return address or
dereference negative address. If such predicate is unsatisfiable, Sydr falls
back to solving the original security predicate. Furthermore, when additionally
both address and value are symbolic, Sydr reports that write-what-where
condition is possible for this out-of-bounds access.

\subsection{Integer Overflow}

Integer overflow is one of the most common program errors~\cite{cwe-top25}.
However, it occurs quite often in binary code. Thus, Sydr would operate too long
if we check all the cases when integer overflow may happen. Moreover, there are
some situations like hash functions where integer overflow is normal. So, we
highlight only critical parts and verify security predicates for them. Unlike
other security predicates, we separate an error sink from its source. Source is
an instruction where integer overflow may happen. And sink is a place in code
where preceding flaw may lead to critical error.

We solve integer overflow security predicates in error sinks that use
potentially overflowed value, i.e. branches (changing control flow depending on
overflowed value), memory access addresses, and function arguments. This is
especially critical for such functions as \texttt{malloc}, \texttt{memcpy}, and
others~\cite{wang09, demidov17}. We wrap some common standard library functions
and consider all their symbolic arguments to be potential error sinks. For other
function calls we check the first three arguments according to standard calling
convention.

First of all, we check whether instruction is arithmetic and one of its operands
is symbolic. If so, we build security predicates for unsigned (CF) and signed
(OF) integer overflow errors. For most arithmetic instructions security
predicate is true when the corresponding flag is set to 1.

Then we figure out whether error source instruction result is involved in
computation of sink. Firstly, we check whether at least one children of sink AST
matches the source AST (overflowed arithmetic instruction result). If so, sink
contains potentially overflowed value in its computations.

\begin{algorithm}[t]
  \caption{Signedness detection via backward slicing.}
  \textbf{Input:} $sink$~-- sink AST node containing overflowed value,
  $call\_stack$~-- call stack on error sink, $\Pi$~-- path predicate (path
  constraints prior to the error sink).  \\
  \textbf{Output:} Return True if type is signed, False if unsigned, None if
  cannot be detected.
  \begin{algorithmic}
    \State $vars \gets used\_variables(sink)$
    \Comment{slicing variables}
    \State \LeftComment{0}{Iterate backward over path constraints.}
    \ForAll {$c \in reversed(\Pi)$}
      \State \LeftComment{0}{Check whether constraint uses sink node $vars$.}
      \If {$vars \cap used\_variables(c) \neq \varnothing$}
        \State \LeftComment{0}{Get instruction corresponding to constraint $c$.}
        \State $inst \gets instruction(c)$
        \State \LeftComment{0}{Get call site of function containing $inst$.}
        \State $s \gets callsite(function(inst))$
        \If {$inst$ is branch \AND $s \in call\_stack$}
          \If {$inst$ is signed} \Comment{js/jns/jg/jge/jl/jle}
            \State \textbf{return} \textbf{True}
          \EndIf
          \If {$inst$ is unsigned} \Comment{ja/jae/jb/jbe}
            \State \textbf{return} \textbf{False}
          \EndIf
          \State \LeftComment{0}{Signedness may be ambiguous (e.g. jz).}
        \EndIf
      \EndIf
    \EndFor
    \State \textbf{return} \textbf{None}
  \end{algorithmic}
  \label{alg:slicing}
\end{algorithm}

Afterwards, we identify the signedness of arithmetic operation. So, we could
select one of signed or unsigned integer overflow security predicates. It is
important because the lack of signedness knowledge leads to many false
positives. We propose Algorithm~\ref{alg:slicing} to detect signedness via
backward slicing~\cite{weiser84}. We iterate backwards starting from the last
branch in path predicate (at the moment of error sink). We locate the first
branch that we can learn signedness from~\cite{wang09}. For example, \texttt{JL}
branch instruction tells us that value is signed. Moreover, the branch must use
at least one symbolic variable from sink AST and its call site have to be in
current call stack, i.e. function containing this branch should be in current
call stack.

Furthermore, we can detect signedness if symbolic value is obtained with
\texttt{strto*l} functions. For instance, \texttt{strtol} is used for signed
value and \texttt{strtoul} for unsigned value.

Finally, when we get information about the signedness (or do not), we check
whether security predicates are satisfiable. If we know the signedness, we check
only one specific security predicate for this signedness. We slice all relevant
constraints from the path predicate at sink point and conjunct them with the
security predicate. We report signed or unsigned integer overflow and generate
corresponding input when the resulting predicate is satisfiable.

If we couldn't retrieve the signedness, then we query whether the integer overflow of both
kinds may happen at the same time. In other words, may some single input cause
unsigned and signed integer overflow errors simultaneously. So, we conjunct
security predicates of both signednesses and sliced path predicate. If the
resulting predicate is true, we report both signed and unsigned overflows and
generate a single corresponding input. Otherwise, we check security predicates
separately. If both security predicates are separately true, we print two
warnings and save two inputs for each of them. We do not report errors when
single signed or unsigned integer overflow is possible without knowing the
signedness since it may lead to false positives.

As we said before, it is crucial when potential sink is argument of functions
like \texttt{malloc}, \texttt{calloc}, \texttt{memcpy}, etc. Thereby, we do not
only check for ordinary integer overflow but we reveal whether even more
dangerous consequences might appear. For allocation functions we examine whether
the size can be overflowed such that it would be a non-zero value smaller than
it was during the concrete execution. In particular, it may further lead to
buffer overflow. For copying functions we investigate whether the size can be
overflowed such that it would be greater than its value during the concrete
execution. For such cases we build additional strong preconditions that are
conjuncted with security predicate. If conjunction with precondition is
unsatisfiable, we fall back to solving the original security predicate.

Let us consider the example of program with potential integer overflow that may
lead to buffer overflow on 32-bit architecture:
\begin{lstlisting}[language=C, basicstyle=\small\ttfamily, numbers=left,
                   xleftmargin=2em,
                   caption={Integer overflow in \texttt{malloc} size (32-bit).},
                   captionpos=b,
                   label=lst:malloc]
int main() {
    int size;
    fscanf(stdin, "%d", &size);
    if (size <= 0) return 1;
    size_t i;
    int *p = malloc(size * sizeof(int));
    if (p == NULL) return 1;
    for (i = 0; i < (size_t)size; i++) {
        p[i] = 0;
    }
    printf("%d\n", p[0]);
    free(p);
}
\end{lstlisting}
We run this program under Sydr with standard input +00000000002 that is enough
to generate 32-bit positive and negative numbers. The program gets the
\texttt{size} of the array to allocate via \texttt{strtol} function called from
\texttt{scanf}. String to integer conversion function semantics are applied to
obtain symbolic formula for \texttt{size}. The array \texttt{size} is multiplied
by \texttt{sizeof(int)} which is a potential error source. We construct signed
integer overflow security predicate for the multiplication (signedness is
identified from \texttt{strtol} call). The potential error sink is in
\texttt{malloc} function argument that uses the multiplied value. Moreover, we
conjunct the security predicate with strong precondition that $size *
sizeof(int)$ should be lower than original $2 * 4 = 8$. This precondition helps
to bypass check in line~7, i.e. we shouldn't allocate more than total memory
available on the machine. Then we conjunct security predicate, precondition, and
sliced path predicate. The obtained predicate
$(long\ long) size * sizeof(int) \neq (int) size * sizeof(int)\ \land\ (unsigned) size * sizeof(int) < 8\ \land\ size > 0$
is passed to solver. The solver reports it's satisfiable and
returns a corresponding model. Sydr generates new standard input value
+01073741825 that triggers integer overflow in line~6. And \texttt{malloc}
allocates only $(unsigned) 1073741825 * sizeof(int) = 4$ bytes because of
wraparound. Thus, in line~9 we get buffer overflow since cycle iterates over
1073741825 array elements when its real size is 1.

\subsubsection{Error Source Pitfalls}

We build security predicates for \texttt{SHL} and \texttt{SAL} instructions
differently because \texttt{CF} flag in these instructions contains the value of
the last bit shifted out of the destination operand and \texttt{OF} flag is
affected only for 1-bit shifts. So, \texttt{CF} and \texttt{OF} flags do not
represent the fact that overflow occurred. Therefore, unsigned integer overflow
security predicate is true when at least one of significant bits was shifted
from destination operand. Signed integer overflow occurs when both zeros and
ones were shifted.

On 32-bit architecture \texttt{int64\_t} addition and subtraction are
implemented as a sequence of \texttt{ADD/ADC} or \texttt{SUB/SBB} instructions.
Therefore, we build the security predicate only for the last \texttt{ADC/SBB} instruction
in sequence.

Moreover, some compiler optimizations replace subtraction of \texttt{int64\_t}
and \texttt{unsigned int} types with addition, e.g. replace \texttt{sub eax, 1}
with \texttt{add eax, 0xffffffff}~\cite{wang09}. When we see an addition with
negative constant, we construct a security predicate just like for subtraction
with positive value.

\subsubsection{Error Sink Pitfalls}

We intend to correctly handle integer promotion. If arithmetic operations on
types smaller than \texttt{int} occur, their operands get extended to bigger
types. For instance, two \texttt{char} operands would be extended to
\texttt{int} and addition would compute 32-bit values instead of 8-bit. Thus,
overflow can never happen in the moment of addition because 32-bit can fit any
result of 8-bit arithmetics. The actual error will be on the step when we
truncate the result to the smaller type back (\texttt{char a, b, c; c = a +
b;}). In terms of SMT~\cite{smt-lib}, the addition result is extracted to fit in
8 bits. Therefore, we search for all extraction nodes in the sink AST that
extract from the source AST. Then we reconstruct new security predicates for
smaller operands with sizes equal to extraction node size.

\section{Implementation}
\label{sec:implementation}

We implement the proposed function semantics and security predicates in our
dynamic symbolic execution tool Sydr~\cite{vishnyakov20}. Sydr utilizes
Triton~\cite{saudel15} to construct symbolic formulas, maintain symbolic
registers and memory states, and build path predicate. The obtained formulas are
solved by Z3~\cite{demoura08}.

Sydr is separated in two processes: concrete and symbolic executors. Concrete
executor runs target program under dynamic binary instrumentation framework
DynamoRIO~\cite{bruening04} and sends events to symbolic executor via shared
memory. These events contain all necessary instructions, registers, and memory
values to perform concolic execution. We handle function semantics the similar
way. DynamoRIO wraps library functions. So, we send event containing function
name, its arguments, and return value for each function mentioned in
Section~\ref{sec:func}. Then we handle this event on symbolic executor side and
apply appropriate symbolic semantics. Symbolic executor updates the shadow heap
every time it receives an event for heap library functions. The shadow stack is
renewed when call/ret instructions are received.

We use xxHash~\cite{xxhash} fast hash algorithm in order to skip already
discovered program errors. So, we update bitmap with all satisfiable security
predicates and don't solve them again. The bit index in this bitmap corresponds
to hash over error type, source, and destination addresses. Moreover, we
remember path constraint index when security predicate is unsatisfiable. Thus,
we can skip solving predicates with greater indexes because they are going to be
unsatisfiable too.

The implementation supports x86 (32 and 64 bit) architecture. However, the
proposed methods are not limited to x86 and may later be implemented for other
architectures.

\section{Evaluation}
\label{sec:evaluation}

\subsection{Function Semantics}

\begin{table*}[t]
\caption{Function Semantics Benchmarking}
\begin{center}
\scriptsize
\begin{tabular}{l r >{\columncolor[gray]{0.9}}r | r >{\columncolor[gray]{0.9}}r | r >{\columncolor[gray]{0.9}}r r >{\columncolor[gray]{0.9}}r | r >{\columncolor[gray]{0.9}}r r >{\columncolor[gray]{0.9}}r}
\toprule
&\multicolumn{4}{c |}{\textbf{Path predicate}}&\multicolumn{8}{c}{\textbf{2-hour benchmark}} \\
\textbf{Application}&\multicolumn{2}{c |}{\textbf{Default}}&\multicolumn{2}{c |}{\textbf{Function Semantics}}&\multicolumn{4}{c |}{\textbf{Default}}&\multicolumn{4}{c}{\textbf{Function Semantics}} \\
&\textbf{Branches}&\textbf{Time}&\textbf{Branches}&\textbf{Time}&\textbf{Accuracy}&\textbf{SAT}&\textbf{Queries}&\textbf{Time}&\textbf{Accuracy}&\textbf{SAT}&\textbf{Queries}&\textbf{Time} \\
bzip2recover&5131&6s&5131&6s&100\%&2101&5131&47m35s&100\%&2101&5131&45m38s \\
cjpeg&8008&19s&6992&18s&100\%&50&2656&120m&100\%&50&3750&120m \\
faad&470585&21m&466697&15m52s&97.11\%&1974&3072&120m&98.91\%&1560&2414&120m \\
foo2lava&910737&21m9s&905592&18m20s&87.1\%&31&5998&120m&99.02\%&205&6668&120m \\
hdp&66070&43s&29265&20s&76.69\%&1171&4122&120m&72.22\%&5893&12172&120m \\
jasper&837643&14m47s&771806&10m37s&99.62\%&8457&22538&120m&96.61\%&9528&24472&120m \\
libxml2&53400&40s&8873&12s&51.27\%&1063&18485&120m&82.44\%&1247&8970&5m53s \\
minigzip&8977&1m4s&8977&1m3s&51.47\%&7569&8977&16m16s&51.47\%&7569&8977&16m16s \\
muraster&7102&5s&4453&4s&99.94\%&3304&6041&120m&100\%&360&470&120m \\
pk2bm&3665&2s&658&1s&99.45\%&183&3664&15m55s&100\%&189&657&4m55s \\
pnmhistmap\_pgm&967187&9m21s&967155&9m2s&99.99\%&19351&28932&120m&100\%&19964&29369&120m \\
pnmhistmap\_ppm&7864&12s&7822&11s&99.07\%&107&7990&27m26s&99.12\%&114&7948&25m31s \\
readelf&62713&41s&13649&10s&87.38\%&1022&9541&120m&85.82\%&2363&6541&120m \\
yices-smt2&19352&17s&10340&11s&73.79\%&4258&16222&120m&70.27\%&5534&11753&11m5s \\
yodl&8329&9s&5340&5s&36.25\%&1153&9403&51m3s&98.26\%&1150&6414&1m50s \\
\bottomrule
\end{tabular}
\label{tbl:functions}
\end{center}
\end{table*}

We evaluate function semantics performance and efficiency on 64-bit Linux
programs~\cite{sydr-benchmark}. We utilize AMD EPYC 7702 (128 cores) server with
256G~RAM for Sydr benchmarking. Table~\ref{tbl:functions} presents the
benchmarking results for default symbolic execution and one with function
semantics applied. Sydr symbolically interprets one execution trace for each
program.

First of all, we measure path predicate construction time and number of
discovered symbolic branches for each application. As we can see, both measured
metrics decrease when modeling function semantics. Thus, we win more time for
exploring new paths via branch inversion. The fewer symbolic branches we have
the less overconstraint we achieve in formulas. Moreover, we do not invert
internal branches in wrapped library functions and model them in a single (or
few) formula. So, we are able to explore more program logic states from one
execution trace. For instance, we can make strings equal during one run in
\texttt{memcmp} function.

Secondly, we run Sydr to invert branches in one solving thread. We limit Sydr execution time with 2
hours. Each query solving time is limited to 10 seconds. In our experiment Sydr
inverts branches in direct order (from first to last in path predicate).
Finally, we measure the accuracy~\cite{vishnyakov20} of branch inversion because
not every satisfiable (SAT) solver query actually inverts the target branch.
\textit{Accuracy} is the percent of satisfiable solver queries that successfully
discover the intended path, i.e. they have the same execution branch trace as
original except the last branch, that should be in inverted direction. For most
applications Sydr shows the best results with function semantics applied. Sydr
either inverts all branches on execution trace faster, or discovers more
(accurate) paths for the same 2-hour limit. The number of discovered paths for
\texttt{faad} and \texttt{muraster} decrease because they contain lots of
symbolic string comparisons. We model each string comparison function with a
single data flow formula. Thus, we skip branches inside these functions. These
branches are useless since they compare single bytes. Instead, we perform full
string comparison. We manually verified that we actually invert complex branches
that use string comparison return values. For example,
\texttt{if~(!strcmp(s,~"abc"))}. However, some solver queries become more
complex because they contain sliced~\cite{vishnyakov20} (conjuncted) string
comparison branch constraints. This fact negatively affects the solver
performance.

\subsection{Security Predicates on Juliet}

Juliet Test Suite for C/C++~\cite{juliet} is a collection of test cases for 118
different CWEs. We adopted Juliet build system to make it suitable for dynamic
analysis. We build only those test cases which read user input, i.e. it can be
modified to cause a program error. Each Juliet source file contains two test
cases: positive and negative. Positive test case has an error while negative
case handles potential flaw via additional checks. These cases are implemented
in \texttt{\_bad} and \texttt{\_good} functions that are called directly from
\texttt{main}. We build these tests in separate binaries: one with potential
flaw and other without errors. Thus, each binary runs a single (positive or
negative) test case. Furthermore, for each binary we build its version with
sanitizers~\cite{song19}. We run Sydr security predicates on binary without
sanitizers. Then we verify that Sydr generates input that actually triggers an
error on sanitizers. We propose Juliet C/C++ Dynamic Test
Suite~\cite{juliet-dynamic} that allows to measure true positive and true
negative rates for a dynamic error detection tool.

It is worth noting that Juliet test cases read symbolic input with
\texttt{fscanf} or \texttt{fgets}+\texttt{atoi}. Both use \texttt{strto*l}
functions for string to integer conversion, which makes function semantics
essential for error detection in Juliet tests.

\begin{table*}[t]
\caption{Juliet Testing Results}
\begin{center}
\scriptsize
\begin{tabular}{ l r | r r >{\columncolor[gray]{0.9}}r|
r r >{\columncolor[gray]{0.9}}r}\toprule
\multirow{2}{*}{\textbf{CWE}} & \multirow{2}{*}{\textbf{P=N}} &
\multicolumn{3}{c|}{\textbf{Textual errors}} &
\multicolumn{3}{c}{\textbf{Sanitizers verification}} \\
&& \textbf{TPR} & \textbf{TNR} & \textbf{ACC} & \textbf{TPR} & \textbf{TNR} & \textbf{ACC} \\
CWE121: Stack Based Buffer Overflow & 188 & 100\% & 100\% & 100\% & 100\% & 100\% & 100\% \\
CWE122: Heap Based Buffer Overflow & 376 & 100\% & 100\% & 100\% & 100\% & 100\% & 100\% \\
CWE124: Buffer Underwrite & 188 & 100\% & 100\% & 100\% & 100\% & 100\% & 100\% \\
CWE126: Buffer Overread & 188 & 100\% & 100\% & 100\% & 100\% & 100\% & 100\% \\
CWE127: Buffer Underread & 188 & 100\% & 100\% & 100\% & 100\% & 100\% & 100\% \\
CWE190: Integer Overflow & 2580 & 99.92\% & 90.89\% & 95.41\% & 98.10\% & 90.89\% & 94.50\% \\
CWE191: Integer Underflow & 1922 & 99.90\% & 91\% & 95.45\% & 97.45\% & 91\% & 94.22\% \\
CWE194: Unexpected Sign Extension & 752 & 100\% & 100\% & 100\% & 100\% & 100\% & 100\% \\
CWE195: Signed to Unsigned Conversation Error & 752 & 99.87\% & 100\% & 99.93\% & 99.87\% & 100\% & 99.93\% \\
CWE369: Divide by Zero & 564 & 66.67\% & 100\% & 83.33\% & 66.67\% & 100\% & 83.33\% \\
CWE680: Integer Overflow to Buffer Overflow & 188 & 100\% & 100\% & 100\% & 100\% & 100\% & 100\% \\
\midrule
\textbf{TOTAL} & 7886 & 97.55\% & 94.83\% & 96.19\% & 96.36\% & 94.83\% & 95.59\% \\
\bottomrule
\end{tabular}
\label{tbl:juliet}
\end{center}
\end{table*}

The evaluation results are presented in Table~\ref{tbl:juliet}. We measure Sydr
on a subset of Juliet CWEs. Out-of-bounds security predicate cover CWE 121, 122,
124, 126, 127, 194, and 195. Integer overflow checker detects errors for CWE
190, 191, and 680. CWE369 is handled by division by zero predicate. We do not
build some CWEs (like CWE476: NULL Pointer Dereference) because they have no
user input. Every covered CWE is built for both 32-bit and 64-bit targets. The
only exception is CWE680 (Integer Overflow to Buffer Overflow) that is built
only for 32-bit. For instance, in Listing~\ref{lst:malloc} input variable
\texttt{size} is a 32-bit integer. Potential integer overflow happens in
\texttt{malloc} function argument. Inputted \texttt{size} is multiplied by
\texttt{sizeof(int)} that has \texttt{size\_t} type. On 64-bit architecture
\texttt{size\_t} is equivalent for \texttt{unsigned long long} and 32-bit value,
sign-extended to 64-bit, cannot cause integer overflow.

Juliet Dynamic testing system evaluates the number of true positive ($TP$) and
true negative ($TN$) cases. Afterwards, it computes true positive rate $TPR =
\frac{TP}{P}$, true negative rate $TNR = \frac{TN}{N}$, and accuracy $ACC =
\frac{TP+TN}{P+N}$ (Juliet has equal number of positive and negative cases
$P=N$). We collect these results for two categories of alarmed errors
(Table~\ref{tbl:juliet}):
\begin{itemize}
  \item Textual errors~-- Sydr reports that test contains an error.
  \item Sanitizers verification~-- generated inputs for errors from previous
    category trigger sanitizers.
\end{itemize}
We describe evaluation process below.

Firstly, we collect paths for test case binaries and select appropriate inputs.
We crafted inputs for \texttt{int64\_t}, \texttt{int}, \texttt{short}, and
\texttt{char} input data types. For instance, we pass +00000000002 to standard
input for \texttt{int} type. The plus sign ('+') allows Sydr to change the
number sign. Extra zeros are required to let Sydr pick up big numbers because
DSE cannot increase the number of symbolic input bytes. However, we need
additional first extra zero right after the plus sign since \texttt{scanf}
function makes first zero concrete:
\begin{lstlisting}[language=C, basicstyle=\small\ttfamily,
                   xleftmargin=2em]
if (buf[i] == '0') buf[i] = '0';
\end{lstlisting}

Secondly, we run Sydr on each test case with corresponding inputs. If Sydr
reports an error for positive test, then the result is classified as true
positive. If Sydr alarms on negative test, then it is a false positive. We
suppose that the result is true negative when Sydr reports no errors for
negative test. The false negative result happens when Sydr does not alarm an
error on a positive test.

Finally, we verify generated inputs on previous step with sanitizers. Sydr
generates one or multiple inputs that should reproduce a true positive error.
For instance, there may be multiple errors for nested function sinks. We run all
generated inputs for a single test on sanitizers. If at least one input leads to
sanitizers error detection, it stays as true positive. Otherwise, we change it
to false negative after sanitizers verification. False positive and true
negative results stay the same.

Table~\ref{tbl:juliet} contains measured $TPR$, $TNR$, $ACC$ for textually
alarmed errors and their verification on sanitizers. We conclude that Sydr has
95.59\% overall accuracy for 11 CWEs from Juliet. Sydr completely covers test
suites for CWE 121, 122, 124, 126, 127, 194, and 680. Security predicates miss
some division by zero (CWE369) errors because Sydr is based on
Triton symbolic engine~\cite{saudel15} that doesn't support floating point. As
you can see, integer overflow and underflow (CWE190/191) have some false
positive and false negative errors. On 32-bit architecture \texttt{int64\_t}
type produces large number arithmetics that our security predicates support for
addition and subtraction. However, we haven't implemented multiplication yet.
Other problem is caused by subtractions of \texttt{int64\_t} on 32-bit
architecture. Some compiler optimizations replace subtraction of
\texttt{int64\_t} types with addition, e.g. replace \texttt{sub eax, 1} with
\texttt{add eax, 0xffffffff}. We support this for regular arithmetics but not
for large arithmetics. Moreover, there are some wrong results in test cases
containing squaring of \texttt{short} type that need further investigation.

Sydr evaluation artifacts are available in Juliet Dynamic
repository~\cite{juliet-dynamic}. Thus, one can run the provided script to
reproduce the results.

\section{Conclusion}
\label{sec:conclusion}

We propose security predicates that enable bug detection in our dynamic
symbolic execution tool Sydr~\cite{vishnyakov20}. In particular, Sydr is able to
report null pointer dereference, division by zero, out-of-bounds access, and
integer overflow errors. Unlike static analysis tools, Sydr also generates new
input data that reproduce detected errors. We present function semantics
modeling for common C/C++ standard library functions that assists bug detection,
increases path exploration speed, and reduces the number of overconstrained
formulas. Last but not least, we introduce Juliet Dynamic testing
system~\cite{juliet-dynamic} that allows to measure dynamic bug detection tools
accuracy on Juliet test suite~\cite{juliet}. Sydr achieves 95.59\% overall
accuracy for 11 CWEs from Juliet.

\printbibliography

\end{document}